\documentclass[aps, prc, twocolumn, floatfix, nofootinbib, preprintnumbers, superscriptaddress, amsmath, amssymb, longbibliography]{revtex4-2}
\usepackage{graphicx}
\usepackage{bbm}
\usepackage{slashed}
\usepackage{dcolumn}
\usepackage{siunitx}
\usepackage{multirow}
\usepackage{hhline}
\usepackage{enumitem}
\usepackage{mathrsfs}
\usepackage{xcolor}
\usepackage{array}
\usepackage{ulem}
\usepackage{float}
\usepackage{xspace}

\newcommand{\NNLO}{\ensuremath{{\rm N}{}^2{\rm LO}}\xspace}
\newcommand{\NNNLO}{\ensuremath{{\rm N}{}^3{\rm LO}}\xspace}
\newcommand{\NNNNLO}{\ensuremath{{\rm N}{}^4{\rm LO}}\xspace}
\newcommand{\NkLO}[1]{\ensuremath{\mathrm{N}^{#1}\mathrm{LO}}\xspace}

{}  

\DeclareMathOperator{\pr}{pr} 

\graphicspath{{Figures/}}

\definecolor{bobcatgreen}{rgb}{0.05,0.31,0.25}

\begin{document}

\title{Analyzing rotational bands in odd-mass nuclei\\using Effective Field Theory and Bayesian methods}

\author{I.~K.~Alnamlah}
\email{ia151916@ohio.edu}
\affiliation{Institute of Nuclear and Particle Physics and Department of Physics and Astronomy, Ohio University, Athens, OH 45701,USA}
\affiliation{Department of Physics and Astronomy, King Saud University, Riyadh 11451, Saudi Arabia}

\author{E.~A.~Coello P\'erez}
\email{coelloperez1@llnl.gov}
\affiliation{Lawrence Livermore National Laboratory, Livermore, CA 94550, USA}

\author{D.~R. Phillips}
\email[]{phillid1@ohio.edu}
\affiliation{Institute of Nuclear and Particle Physics and Department of Physics and Astronomy, Ohio University, Athens, OH 45701,USA}

\date{\today}

\begin{abstract}
We recently developed an Effective Field Theory (EFT) for rotational bands in odd-mass nuclei. Here we use EFT expressions to perform a Bayesian analysis of data on the rotational energy levels of $^{99}$Tc, ${}^{155,157}$Gd, ${}^{159}$Dy, ${}^{167, 169}$Er, ${}^{167, 169}$Tm, ${}^{183}$W, ${}^{235}$U and ${}^{239}$Pu. The error model in our Bayesian analysis includes both experimental and EFT truncation uncertainties. It also accounts for the fact that low-energy constants (LECs) at even and odd orders are expected to have different sizes. We use Markov Chain Monte Carlo (MCMC) sampling to explore the joint posterior of the EFT and error-model parameters and show both the LECs and the expansion parameter, $Q$, can be reliably determined. We extract the LECs up to fourth order in the EFT and find that, provided we correctly account for EFT truncation errors in our likelihood, results for lower-order LECs are stable as we go to higher orders. LEC results are also stable with respect to the addition of higher-energy data. We extract the expansion parameter for all the nuclei listed above and find a clear correlation between the extracted and the expected $Q$ based on the single-particle and vibrational energy scales. However, the $Q$ that actually determines the convergence of the EFT expansion is markedly smaller than would be naively expected based on those scales. 
\end{abstract}
\maketitle
\section{Introduction}

Rotational bands are ubiquitous in the spectra of medium-mass and heavy nuclei. As has been known for seventy years~\cite{Bohr:1951zz}, they emerge in a description of the nucleus as a nearly rigid axially-symmetric rotor~\cite{Rowe}. For even-even nuclei the simplest rotational bands consist of $0^+$, $2^+$, \ldots states and their energies are described by an expansion in powers of $I(I+1)$~\cite{Papenbrock:2010,CoelloPerez:2015}. This behavior has recently been obtained in {\it ab initio} calculations of the Be isotope chain~\cite{Caprio:2013,Maris:2014,Jansen:2015,Caprio:2019yxh,McCoy:2020xhp} and $^{34}$Mg~\cite{Hagen:2022tqp}. 

Odd-mass neighbors of a rotor nucleus can then be understood as a fermion coupled to the rotor. The fermion dynamics is simpler in the intrinsic frame in which the nucleus is not rotating, but this frame is non-inertial, so solving the problem there induces a Coriolis force proportional to $\vec{j} \cdot \vec{I}$, the dot product of the single-fermion angular momentum and the total angular momentum of the fermion-rotor system. When combined with other mechanisms, such as excitation of the fermion to higher-single particle states and the fermion disturbing the rotor, this induces a string of terms in the energy-level formula~\cite{BohrMottelson}. Odd powers of $I$  appear, and produce staggering between adjacent levels. Which powers of $I$ are present depends on the value of the quantum number, $K$, the projection of the fermion angular momentum on the rotor axis. For $K=1/2$ bands the energy-level formula is: 
\begin{eqnarray}
&& E(I)=A_K I(I+1) + E_K
 + A_{1} (-1)^{I+1/2} (I+1/2) \nonumber\\
 && \quad +  B_{1} I(I+1) (-1)^{I+1/2} (I+\tfrac{1}{2}) 
+ B_K [I(I+1)]^2
\label{eq:energies}
\end{eqnarray}
where $A_K$, $E_K$, $A_1$, $B_1$, and $B_K$ are parameters, related to rotor properties and single-particle matrix elements, that need to be either derived from a microscopic model or estimated from data.

Over the years a number of models have had success describing this pattern from  underlying density functional theory~\cite{Dudek:1981,Cwiok:1980,Afanasjev:2013,Zhang:2020} or shell-model~\cite{Inglis:1954,Velazquez:1999,Liu:2004,Zhang:2020} dynamics. The models also predict specific values for the coefficients that appear in Eq.~(\ref{eq:energies}). 
In Ref.~\cite{Alnamlah:2020} we took a different approach, organizing the formula (\ref{eq:energies}) as an effective field theory (EFT) expansion in powers of the small parameter $Q \equiv 1/I_{br}$, with $I_{br}$ the spin of the nuclear state at which dynamical effects associated with single-particle and/or vibrational degrees of freedom cause the polynomial expansion in powers of $I$ to break down. This description of rotational bands in odd-mass nuclei builds on the successful EFT developed for even-even nuclei in Refs.~\cite{Papenbrock:2010,CoelloPerez:2015}. Other efforts to develop an EFT for these rotational bands can be found in Refs.~\cite{Papenbrock:2020zhh,Chen:2020qbf}. 

In the odd-mass rotor EFT, Eq.~(\ref{eq:energies}) is the next-to-next-to-next-to-next-to leading order (\NkLO{4}) result for the energies, and the first corrections to it are $\mathcal{O}(E Q^4)$. The EFT analysis of Eq.~(\ref{eq:energies}) organizes it in terms of increasingly accurate predictions: the N$^k$LO energy-level formula has accuracy $\mathcal{O}(E Q^{k})$. All short-distance/high-energy physical mechanisms that affect the energies up to that accuracy are subsumed into the parameters or low-energy constants (LECs) that multiply the $I$-dependent terms in Eq.~(\ref{eq:energies}).  In Ref.~\cite{Alnamlah:2020} we determined these LECs by fitting the lowest levels in the different rotational bands we analyzed. However, this runs the risk of fine-tuning the values of the LECs to those levels, and it does not provide uncertainty estimates for them. Better parameter estimation would use all the data available on a particular band, and account for the $\mathcal{O}(E Q^{k})$ truncation uncertainty present at order N$^k$LO~\cite{Schindler:2009,Furnstahl:2014}. 

Bayesian methods for EFT parameter estimation do just that~\cite{Schindler:2009,Wesolowski:2016,Wesolowski:2019,Wesolowski:2021}. Reference~\cite{Wesolowski:2019} showed that the effect of neglected terms in the EFT expansion could be included in the error model by modifying the likelihood so that the covariance matrix that appears there includes both  experimental uncertainties and EFT truncation errors. More recently, Ref.~\cite{Wesolowski:2021} showed that MCMC sampling of that likelihood enabled the simultaneous determination of the LECs and the parameters of the error model, i.e., the value of $Q$ and the typical size of the ``order 1'' dimensionless coefficients that appear in the EFT expansion.

In this work we apply the EFT parameter estimation technology developed in Refs.~\cite{Schindler:2009,Wesolowski:2016,Wesolowski:2019,Wesolowski:2021} to the problem of rotational bands in odd-mass nuclei. We consider $K=1/2$ bands in $^{99}$Tc, ${}^{167, 169}$Er, ${}^{167, 169}$Tm, ${}^{183}$W, ${}^{235}$U and ${}^{239}$Pu as well as $K=3/2$ bands in ${}^{155,157}$Gd and ${}^{159}$Dy. 
Section~\ref{sec:rotorfermionEFT} summarizes the elements of the EFT that are relevant for this paper.  Section~\ref{sec:Bayesianmodel} then develops the Bayesian statistical model we use to analyze data on rotational bands. We first write down the likelihood that includes both experimental and theory uncertainties, and then explain how we use known information on the expected size of the LECs and the expansion parameter to set priors. A novel feature of this work, compared to earlier Bayesian EFT parameter-estimation studies, is that our statistical model incorporates the possibility that the LECs at even and odd orders have different typical sizes. This reflects the physics of odd-order LECs that are associated with matrix elements of the fermion spin, while even-order LECs contain a combination of effects from the rotor and the fermion. Section~\ref{sec:sampler} contains details of our Markov Chain Monte Carlo sampler, and then Sec.~\ref{sec:results} presents the results for LECs and the expansion parameter, $Q$, that we obtain from sampling the Bayesian posterior. We conclude in Sec.~\ref{sec:conclusion}. All the results and figures generated from this work can be reproduced using publicly available Jupyter notebooks~\cite{notebooks}. 

\section{Rotational EFT Background}

\label{sec:rotorfermionEFT}

Here we summarize the results of the EFT for rotational bands in odd-mass nuclei that was developed up to fourth order in the angular velocity of the system in Ref.~\cite{Alnamlah:2020}. This theory constructs the Lagrangian of the particle-rotor system using its angular velocity and the angular momentum of the unpaired fermion, $\vec{j}$, as building blocks. The resulting Lagrangian corrects that of a rigid rotor with contributions arranged as a series in powers of a small expansion parameter, $Q$, according to a power-counting scheme that counts powers of the system's angular velocity. Naively, we expect $Q$ to be of order $E_{\rm rot}/ E_{\rm high}$, where $E_{\rm rot}$ is the energy scale at which rotational excitation take place and $E_{\rm high}$ is the scale of high-energy physics not explicitly taken into account by the EFT.
At leading order (LO), the energy of a rotational band on top of a bandhead with spin $K$ is
\begin{equation}
\label{Energy_LO}
E_{\rm LO}(I,K) =A_{\rm rot} I(I+1) + E_K,
\end{equation}
where $I$ is the spin of the rotational state (or, equivalently, the total angular momentum of the fermion-rotor system), and $A_{\rm rot}$ and $E_K$ are LECs that must be fitted to experimental data. $A_{\rm rot}$ is determined by the moment of inertia of the even-even nucleus (the rotor) to which the unpaired fermion is coupled.

At next-to-leading order (NLO) rotational bands with $K=1/2$ are affected by a term that takes the same $\vec{j} \cdot \vec{I}$ form as the Coriolis force. This produces:
\begin{equation}
\label{Energy_NLO}
\begin{aligned}
E_{\rm NLO}(I,K) =& A_{\rm rot} I(I+1) + E_K\\
& + A_{1} (-1)^{I+1/2} \left(I+\tfrac{1}{2}\right) \delta^K_{1/2}.
\end{aligned}
\end{equation}
The LEC $A_1$ is expected to be of order $A_{\rm rot}$ times a sum of matrix elements involving the fermion's total angular momentum operator (for details see Ref.~\cite{Alnamlah:2020}). From previous studies we see that $A_1/A_{\rm rot}<1$. This correction, sometimes called the signature term, causes staggering between adjacent states in $K=1/2$ bands.

The energy of a rotational band at next-to-next-to-leading order (\NNLO) is 
\begin{equation}
\begin{aligned}
E_{\NNLO}(I,K) =& A_K I(I+1) + E_K\\
&+ A_{1} (-1)^{I+1/2} \left(I+\tfrac{1}{2}\right) \delta^K_{1/2}.
\end{aligned}
\end{equation}
The term proportional to $A_K$ combines the LO term proportional to $A_{\rm rot}$ and corrections entering at this order with the same spin dependence. From our power counting we expect the shift $\Delta A=A_{\rm rot}-A_K$ to be of order $A_{\rm rot} Q$. In contrast to $A_{\rm rot}$, $A_K$ is band dependent and so should be fitted to data on the rotational band of interest. 

The \NNNLO corrections to the energy of a rotational band are both $\sim I^3$ for $I \gg 1$, but take a different form in the $K=1/2$ and $K=3/2$ bands:
\begin{equation}
\begin{aligned}
\Delta E_{\NNNLO}(&I,K)\\
= B_{1}& (-1)^{I+1/2} \left(I+\tfrac{1}{2}\right)I(I+1)  \delta^K_{1/2} \\
+& A_{3} (-1)^{I+3/2} \left(I+\tfrac{1}{2}\right)\left(I-\tfrac{1}{2}\right)\left(I+\tfrac{3}{2}\right) \delta^K_{3/2}.
\end{aligned}
\end{equation}
with $B_1$ and $ A_3$ expected to be of order $A_1 Q^2$.
Last, at \NNNNLO we have the additional term:
\begin{equation}
\Delta E_{\NNNNLO}(I,K) = B_K [I(I+1)]^2.
\end{equation}
with $B_K$ expected to be of order $A_{\rm rot} Q^3$.

This pattern continues: at odd orders we add terms that correct the staggering term and have LECs of order $A_1Q^{n-1}$, while the even-order terms provide the overall trend with $I$ and have LECs of order $A_{\rm rot}Q^{n-1}$. (In both cases $n$ is the order of our expansion.) This difference in the expected sizes of odd and even LECs comes from the physics. Odd-order LECs are associated with operators in the effective Lagrangian that couple rotor and fermionic degrees of freedom, while even-order LECs encode both rotor-fermion interactions and effects coming from the non-rigidity of the rotor itself.

In what follows we denote the LECs $A_1$, $\Delta A$, $B_1$, and $B_K$ generically as $\{a_n:n=1,\ldots,k\} \equiv {\bf a}_k$, where $k$ is the order of the EFT calculation. (In the case of $K=3/2$ bands the set is $\Delta A$, $A_3$, and $B_K$, and $a_1=0$.)
We then divide the $n$th-order LEC, $a_n$, by the the reference scale and the power of the expansion parameter assigned to it by the EFT power counting, i.e., construct:
\begin{equation}
    c_n=\frac{a_n}{A_{\rm rot} Q^{n-1}}.
    \label{eq:cs}
\end{equation}
We expect these coefficients $c_n$ to be of order one, i.e., they should be natural coefficients. However, because sets of odd and even natural coefficients seem to have different sizes we will assume the even and odd $c_n$'s are drawn from two different distributions with different characteristic sizes that we denote by $\bar{c}_{even}$ and $\bar{c}_{odd}$. 

\section{Building the Bayesian Model}

\label{sec:Bayesianmodel}

\subsection{Building the Posterior}

Our goal in this analysis is to use the information on the expected size of LECs to stablize the extraction of their values as we add more levels to the analysis, or as we use energy-level formulae computed at different EFT orders. At the same time, we want to estimate the expansion parameter, $Q$, of the theory, as well as the characteristic sizes for even and odd coefficients, $\bar{c}_{even}$ and $\bar{c}_{odd}$. 

We want to obtain the posterior distribution for all the LECs that appear at order $k$, a set we collectively denote by ${\bf a}_k$. Here we will obtain the joint posterior pdf of ${\bf a}_k$, the expansion parameter $Q$, and the characteristic sizes. To do this we follow the successful endeavor by the BUQEYE collaboration in Refs.~\cite{Wesolowski:2016, Wesolowski:2019, Wesolowski:2021}, and write the posterior, given experimental data, $\vec{y}_{exp}$, and prior information on the model, $P*$, as
\begin{equation}
\begin{aligned}
    \pr({\bf a}_k,Q,\bar{c}_{even},\bar{c}_{odd}|&\vec{y}_{exp},P*)\\
    =&\pr({\bf a}_k|Q,\bar{c}_{even},\bar{c}_{odd},\vec{y}_{exp},P*)\\
    \times&\pr(Q|\bar{c}_{even},\bar{c}_{odd},\vec{y}_{exp},P*)\\
    \times&\pr(\bar{c}_{even}|\bar{c}_{odd},\vec{y}_{exp},P*)\\
    \times&\pr(\bar{c}_{odd}|\vec{y}_{exp},P*).
    \label{eq:master}
\end{aligned}
\end{equation}
Marginalization of this posterior distribution over $Q$, $\bar{c}_{even}$ and $\bar{c}_{odd}$ yields the posterior distribution for ${\bf a}_k$. Other marginalizations can be carried out to obtain posteriors for $Q$, $\bar{c}_{even}$ and $\bar{c}_{odd}$.

This joint posterior distribution tells us the probability of the LECs and the error model parameters given experimental data. We could use this posterior distribution to get other quantities or observables, such as the energy of a particular rotational level, which depend on the LECs or the error model parameters. These are now represented by distributions and not single numbers. Their distributions are called posterior predictive distributions (PPD). We write the PPD of an observable $\mathcal{O}$ as
\begin{equation}
\label{PPD}
    \pr(\mathcal{O}|\vec{y}_{exp},P*)=\int d\vec{\theta}\delta(\mathcal{O}-\mathcal{O}(\vec{\theta})) \pr(\vec{\theta}|\vec{y}_{exp},P*)
\end{equation}
where $\vec{\theta}$ represents the LECs and the error model parameters. Calculating the observable at each point in the parameter space $\vec{\theta}$ and then integrating over the parameters $\vec{\theta}$ allows one to carefully account for correlations between the parameters.

Using Bayes' theorem, we can express the posterior (\ref{eq:master}) as
\begin{equation}
\label{posterior}
\begin{aligned}
    \pr({\bf a}_k,Q,\bar{c}_{even},\bar{c}_{odd}|&\vec{y}_{exp},P*)\\
    =&\pr(\vec{y}_{exp}|{\bf a}_k,Q,\bar{c}_{even},\bar{c}_{odd},P*)\\
    \times& \pr({\bf a}_k|Q,\bar{c}_{even},\bar{c}_{odd},P*)\\
    \times& \pr(Q|\bar{c}_{even},\bar{c}_{odd},P*)\\
    \times& \pr(\bar{c}_{even}|P*)\pr(\bar{c}_{odd}|P*)\\
    \times& \frac{1}{{\pr(\vec{y}_{exp}|P*)}}.
\end{aligned}
\end{equation}
The terms on the right-hand side of Eq.~(\ref{posterior}) have the following interpretations: 
\begin{enumerate}
    \item $\pr(\vec{y}_{exp}|{\bf a}_k,Q,\bar{c}_{even},\bar{c}_{odd},P*)$ is the likelihood of the experimental data given specific values of both the LECs that appear in the energy formula at order $k$ and the parameters in our error model.
    \item $\pr({\bf a}_k|Q,\bar{c}_{even},\bar{c}_{odd},P*)$ is the prior distribution of the LECs given the parameters encoding the systematic expansion of the EFT.
    \item $\pr(Q|\bar{c}_{even},\bar{c}_{odd},P*)$ is the prior distribution of the expansion parameter given the characteristic sizes of even and odd natural coefficients.
    \item $\pr(\bar{c}_{even}|P*)$ and $\pr(\bar{c}_{odd}|P*)$ are the prior distributions of the even and odd characteristic sizes. (In Eq.~\eqref{posterior} we assume an uncorrelated prior on $\bar{c}_{even}$ and $\bar{c}_{odd}$.)
    \item $\pr(\vec{y}_{exp}|P*)$ is the evidence, which we drop in what follows as it does not depend on the parameters we are interested in extracting and functions only as a normalization constant.
\end{enumerate}
\subsection{Building the Likelihood}
We now build the likelihood function accounting for the expected error between the experimental and theoretical values, for data on $K=1/2$ rotational bands. The corresponding likelihood for $K=3/2$ bands is built analogously. Following~\cite{Wesolowski:2019} we start by writing our observable (the energy of a particular rotation level) at order $k$ as
\begin{equation}
\label{eq:energy}
\begin{aligned}
    E(I) &= A_{\rm rot}I(I+1)\\
    \Bigg\{1 &+ \sum_{n=odd}^{k} c_n Q^{n-1}
    (-1)^{I+1/2}\left(I+\tfrac{1}{2}\right)[I(I+1)]^{(n-3)/2}\\
    &+ \sum_{n=even}^{k} c_n Q^{n-1} [I(I+1)]^{(n-2)/2}\Bigg\}.
\end{aligned}
\end{equation}
We choose the leading-order energy for each level, $A_{\rm rot}I(I+1)$, to be the reference scale $E_{\rm ref}$ for the observable. The dimensionless coefficients $c_n$ (see Eq.~(\ref{eq:cs})) are assumed to be $\mathcal{O}(1)$. The theory error $\vec{\sigma}_{th}$ at any order is due to terms omitted from the summations in Eq.~\eqref{eq:energy}. Its most significant contribution comes from the first omitted term in the EFT expansion. Accounting only for this term yields an estimate for the theory error that is fully correlated across levels if $k+1$ is even, and anticorrelated for adjacent levels if $k+1$ is odd. To account for this correlation or anticorrelation we write the theory covariance matrix as the outer product of a vector representing the theory error, $\Sigma_{th} \equiv \vec{\sigma}_{th} \otimes \vec{\sigma}_{th}$.
The vector $\vec{\sigma}_{th}$ contains the value of the first omitted term for each of the $m$ energy levels that enter the likelihood.
We also account for experimental errors by writing the covariance matrix as
\begin{equation}
\Sigma=\Sigma_{th}+\Sigma_{exp}
\end{equation}
where we take $(\Sigma_{exp})_{ij} \equiv (\vec{\sigma}_{exp})_i^2\delta_{ij}$.
The likelihood function is then 
\begin{equation}
\begin{aligned}
    \pr(\vec{y}_{exp}|&{\bf a}_k,Q,\bar{c}_{even},\bar{c}_{odd},P*)\\
    =& \sqrt{\frac{1}{(2\pi)^m |\Sigma|}} \exp\left( -\frac{1}{2}\vec{r}^T \Sigma^{-1} \vec{r} \right),
\end{aligned}
\end{equation}
where $\vec{r}\equiv\vec{y}_{exp}-\vec{y}_{th}$ is the residual between the central experimental energy for a level and the theory result (\ref{eq:energy}) and 
$m$ is the number of levels included in the likelihood estimation.

We note that since the theory error is the outer product of the theory error with itself, the theory covariance $\Sigma_{th}$ is singular. Including the experimental error solves this singularity problem for the covariance $\Sigma$. However, $\Sigma$ can still become ill-conditioned for higher values of $Q$ if the experimental errors are too small; numerical issues then arise when we try to invert the covariance matrix. 

\begin{figure}[t]
    \centering
    \includegraphics[width=\columnwidth]{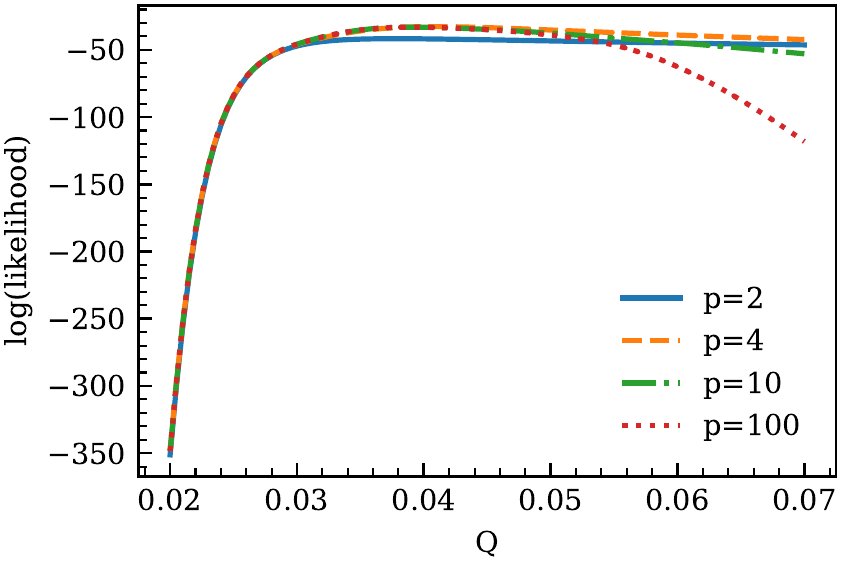}
    \caption{Comparing the Log of the likelihood when accounting for different number of omitted terms, $p$, in the theory error. Apart from $Q$, the parameters that enter the likelihood were chosen to be the median parameters after we have had sampled the posterior distribution for $^{169}$Er.}
    \label{fig:likelihood}
\end{figure}

Including more terms in the estimate for the theoretical error produces a steeper peak in the likelihood function, see Fig.~\ref{fig:likelihood}, which, in turn, restricts the values sampled for $Q$ to a narrower region. Because it precludes the sampler exploring large values of $Q$, this inclusion of more omitted terms in the model of the theoretical error solves the numerical problem of ill-conditioned matrices and gives a more accurate extraction of the LECs and the error-model parameters.

In what follows we estimate the theory error including omitted terms up to a certain cutoff order $k_{max}$. Our theory error estimate for the level with spin $I$ is then
\begin{equation}
\sigma_{th}(I)=A_{\rm rot}\sum_{l=k+1}^{k_{max}}\bar{c}_{even,odd} Q^{l-1} P_l(I),
\label{eq:theoryerror}
\end{equation}
where the $\bar{c}$ that is used here is $\bar{c}_{even}$ for even values of $l$ and $\bar{c}_{odd}$ otherwise. The $I$-dependence of the $l$th term is chosen to match that in Eq.~\eqref{eq:energy}, and is denoted here by $P_l(I)$, a polynomial of power $l$.
We arrange the contributions to the theory error, (\ref{eq:theoryerror}) as the $p$ columns of a $m\times p$ matrix $\sigma_{th}$, where $p=k_{max}-k$ is the number of omitted terms. 
Each column in this matrix then  corresponds to the theory-error structure, while each row corresponds to a different energy level. To obtain $\Sigma_{th}$ we then again take the outer product of $\sigma_{th}$ with itself, i.e., we construct an outer product in our $m$-dimensional data space, while also taking an inner product in order space.
This results in the theory error associated with different orders being added in quadrature, while maintaining the correlation structure of the theory error across the data space.

\subsection{Building the Priors}
\label{priors}
The prior distributions for an order-$n$ LEC is taken to be a Gaussian with mean zero and standard deviation 
\begin{equation}
    \sigma_n= \left\{ \begin{array}{lc}
        A_{\rm rot} \bar{c}_{even} Q^{n-1} & \mbox{if $n$ is even;}\\
        A_{\rm rot} \bar{c}_{odd} Q^{n-1}
        & \mbox {if $n$ is odd.} \end{array} \right.
\label{eq:sdeviation}
\end{equation}
encoding the EFT expectations for the sizes of the LECs arising from the power counting described in Sec.~\ref{sec:rotorfermionEFT}. The standard deviation in Eq.~\eqref{eq:sdeviation} allows the possibility for even and odd LECs to have different typical sizes. Combining the Gaussian priors for the LECs yields
\begin{equation}
\begin{aligned}
    \pr({\bf a}_k|&Q,\bar{c}_{even},\bar{c}_{odd},P*)\\=&\frac{1}{\bar{E} \sqrt{2\pi}} \exp\left(-\frac{E_k^2}{2 \bar{E}^2}\right)
    \prod_{n=1}^k \frac{1}{\sigma_n \sqrt{2\pi}}{\rm exp}\left(-\frac{a_n^2}{2\sigma_n^2}\right).
\end{aligned}
\end{equation}
The LEC $E_K$ is just an energy shift and its size is not determined by the EFT power counting. We set the prior on it to be Gaussian with mean zero and a standard deviation, $\bar{E}$, that is wide enough to capture its value. The value for $\bar{E}$ is determined from the energy of the bandhead and $A_{\rm rot}$ by means of Eq.~(\ref{Energy_NLO}).

We choose not to impose any expectations regarding the size of expansion parameter in the prior for $Q$ and so take it to be flat between two limits:
\begin{equation}
    \pr(Q|\bar{c}_{even},\bar{c}_{odd},P*)\propto
    \begin{cases} 
      1 & Q \in (0,Q_{cut}) \\
      0 & {\rm otherwise}.
   \end{cases}
\end{equation}
Limiting $Q$ from above restricts the sampler from going to high values of $Q$, as they make the covariance matrix ill-conditioned and harder to invert. For all cases we check that the posterior for $Q$ is confined to values well below $Q_{cut}$.

\begin{figure}
    \centering
    \includegraphics[width=\columnwidth]{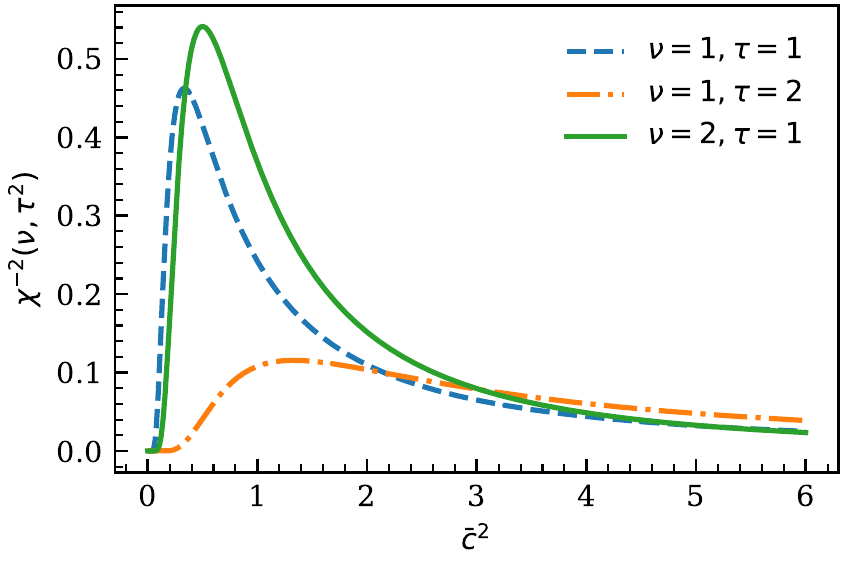}
    \caption{Prior distribution of the size of the dimensionless natural coefficients, $\bar{c}$.}
    \label{fig:prior_c_bars}
\end{figure}

The priors on the characteristic sizes $\bar{c}_{even}$ and $\bar{c}_{odd}$, are taken to be identical scaled-inverse-$\chi^2$ distributions
\begin{equation}
    \pr(\bar{c}_l^2|P*) \propto
    \begin{cases}
    \chi^{-2}(\nu=1,\tau^2=1) & \bar{c}_l^2 \in (0, \bar{c}_{cut}^2) \\
    0 & {\rm otherwise},
    \end{cases}
\end{equation}
where the cutoff $\bar{c}_{cut}$ prevents numerical issues inverting the covariance matrix. The scaled-inverse-$\chi^2$ distribution, given by
\begin{equation}
\chi^{-2}(x; \nu, \tau^2)=
\frac{(\tau^2\nu/2)^{\nu/2}}{\Gamma(\nu/2)}~
\frac{\exp\left[-\frac{\nu \tau^2}{2 x}\right]}{x^{1+\nu/2}},
\end{equation}
is shown for different values of $\nu$ and $\tau$ in Fig.~\ref{fig:prior_c_bars}.
We stress that we chose identical priors for $\bar{c}_{even}$ and $\bar{c}_{odd}$ even though we expect the former to be larger than the latter based on previous analyses of data on rotational bands~\cite{Alnamlah:2020}. We did not want to bias our analysis by imposing this hierarchy on the prior, instead anticipating that it will emerge naturally in the posteriors for those parameters.
 
The scaled-inverse-$\chi^2$ favors small values of $\bar{c}^2$ and has long tails. This allows the sampler to explore higher values of $\bar{c}^2$. The sharp decrease in this distribution for very small values of $\bar{c}^2$ could be a problem for cases where  $\bar{c}_{odd}$ is much smaller than one. This is a concern in some $K=3/2$ bands where we expect smaller odd-order corrections to the leading-order energy than in $K=1/2$ bands.
 \begin{figure*}
    \centering
    \includegraphics[width=2\columnwidth]{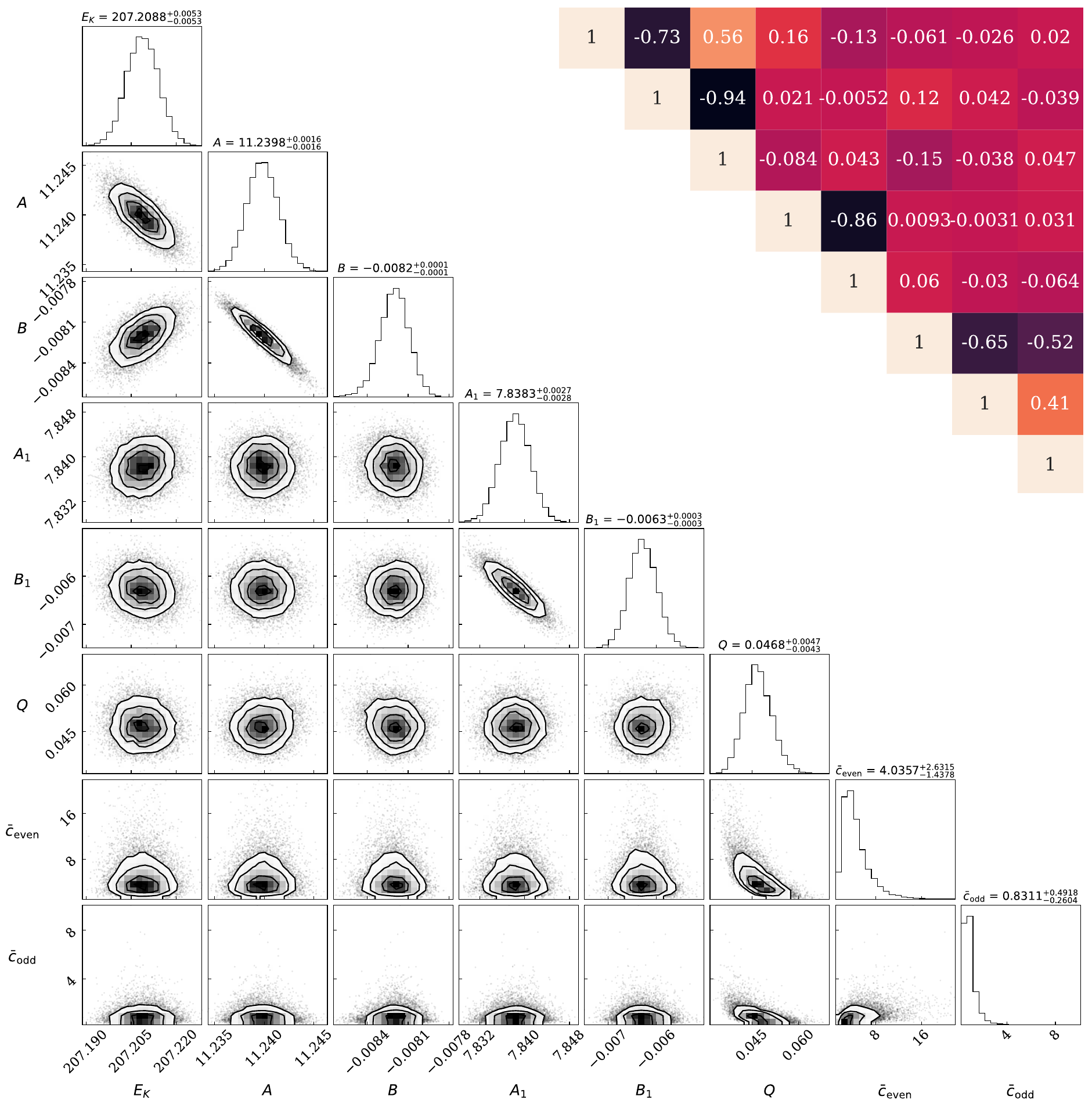}
    \caption{Corner plot for the marginalized distributions of the LECs and the error-model parameters at \NNNNLO for $^{167}$Er including all adopted rotational levels ($I_{max}=16.5$) and accounting for 6 omitted terms in the theory error. The insert in the top right corner show the correlations between posterior parameters. The order of the parameters on the corner plot is the same on the correlations plot. (Here $E_K$ and all the EFT LECs are expressed in keV. The error-model parameters are dimensionless.)}
    \label{fig:corner}
\end{figure*}
\section{Running the Sampler}
\label{sec:sampler}
To sample the posterior distribution in Eq.~\eqref{posterior} we use the Python ensemble sampling toolkit for affine-invariant MCMC (emcee) \cite{emcee}. We run the sampler for each nucleus at a certain EFT order using the $m$ rotational levels from the bandhead up to some $I_{max}$ and accounting for $p$ omitted terms in the theory error. We use 64 walkers to sample the posterior distribution for an initial 10000 steps. We then continue running the sampler with 3000 step increments. After every 3000 steps we calculate the autocorrelation time, $\tau_\alpha$, where $\alpha$ indexes an LEC or an error-model parameter. We declare the sampler to be converged if the sampler meets two criteria. First, the number of steps has to be more than 50 times the highest $\tau_\alpha$.
Second, the change in any of the $\tau_\alpha$'s has to be less than 2\% from its value after the last 3000 step increment.

To get the posterior distributions we discard $2\times {\rm max}(\tau_\alpha)$ steps from the beginning of the chain ({\it burn-in}) and $0.5\times {\rm min}(\tau_\alpha)$ steps in between steps we accept ({\it thinning}).

A sample corner plot of the marginalized distributions of the LECs and the error-model parameters $Q$, $\bar{c}_{even}$ and $\bar{c}_{odd}$, for the case of ${}^{167}$Er is shown in Fig.~\ref{fig:corner}. This figure clearly shows that the posterior distributions for all parameters are fully converged. For this particular case we set $Q_{cut}=0.16$ and $\bar{c}_{cut}=22$ for both $\bar{c}_{even}$ and $\bar{c}_{odd}$. As explained in Sec.~\ref{priors}, the cutoffs on $Q$ and the characteristic sizes prevent the covariance matrix from being ill-conditioned. We also ran the sampler for $^{167}$Er at different values of $Q_{cut}$ and $\bar{c}_{cut}$ and found that different choices of these hyperparameters do not result in a significant change in the posterior distributions.

For some cases, namely $^{99}$Tc and $^{183}$W, the posterior distribution of $Q$ was initially at the upper limit of the prior. We then ran into numerical problems when increasing $Q_{cut}$ trying to encompass the entire posterior. This problem was solved by decreasing the number of levels included in the analysis, i.e., decreasing $I_{max}$. It was then possible to increase $Q_{cut}$ without encountering problems with degenerate matrices. This means that for $^{99}$Tc we were only able to extract the LECs and $Q$ at $I_{max}=11.5$. For $^{183}$W we needed to remove two levels from the upper end of the data set for the sampler to be numerically stable.

In Fig.~\ref{fig:corner} we see clear correlations between $E_K$, $A$, and $B$ and also between $A_1$ and $B_1$. (Here we have dropped the subscript $K$ on $A$ and $B$; it is to be understood that all LECs are band dependent.) The correlation coefficients given in the inset in the top-right corner of the figure make the block-diagonal structure of the covariance matrix clear. To a good approximation the correlation matrix can be decomposed into a correlation matrix for even-order LECs, one for odd-order LECs, and one for the error-model parameters.

We note that, as expected, $\bar{c}_{odd}$ is smaller than $\bar{c}_{even}$. Corrections to the energy levels carrying odd powers of $I$ are smaller than those carrying even powers of $I$. This size difference is connected to different physics correcting the effective Lagrangian at even and odd orders.

To see which of the parameters has the narrowest distribution and therefore places the strongest constraint on the posterior distribution, we did a Singular Value Decomposition (SVD) of the Hessian matrix. We found that the eigenvector with the highest eigenvalue, i.e., the parameter combination with smallest absolute error, is made up mostly of the highest-order LEC. This is unsurprising, since that LEC, $B$, is markedly smaller than the others (we note that its relative error is actually larger than that on, e.g., $A_1$). 

We initially found a peculiar correlation between LECs in some cases where the rotational band was built on the ground state of the nucleus we were looking at. There we found the eigenvector with the highest eigenvalue was a very particular linear combination that involved all the LECs. We ultimately traced this correlation to  the fact that the ground state experimental error had been set to zero, and so the combination of LECs that entered the formula for the ground-state energy was very well constrained (theory error is also very small there). This problem was solved by adding a small experimental error to the ground state. We chose it to be equal to the error that the NNDC quotes on the energy of the first excited state.
\section{Results}
\label{sec:results}
\begin{figure}[b!]
    \centering
    \includegraphics[width=\columnwidth]{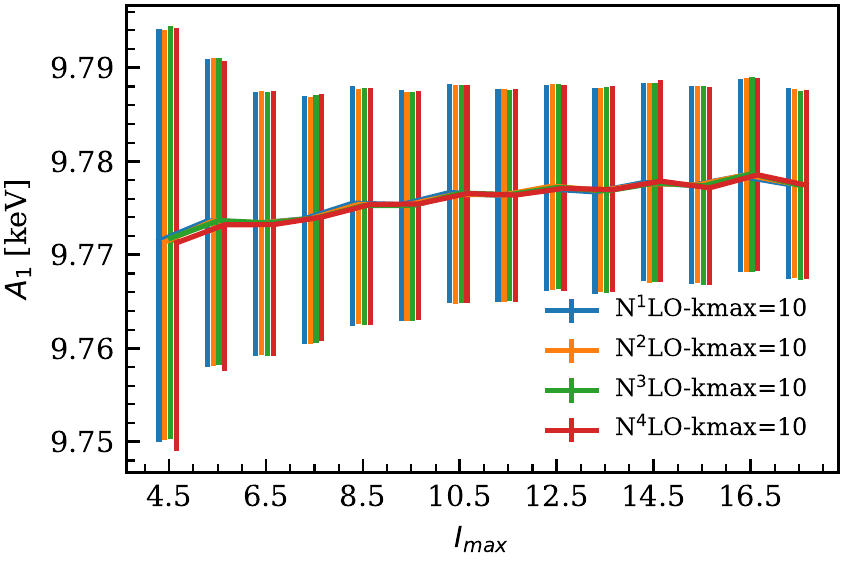}
    \caption{Posteriors for $A_1$ describing $^{169}$Er as a function of $I_{max}$ at different EFT orders. The solid line connects the median values and the error bands encompass the 16th and 84th percentiles of the marginalized distribution.}
    \label{fig:wes_A1_above_orders}
\end{figure}
In this section we show results for our Bayesian analysis of the rotational energy levels in $^{99}$Tc, ${}^{155,157}$Gd, ${}^{159}$Dy, ${}^{167, 169}$Er, ${}^{167, 169}$Tm, ${}^{183}$W, ${}^{235}$U and ${}^{239}$Pu. The experimental data are taken from the National Nuclear Data Center (NNDC)~\cite{Baglin:2016vll,Browne:2017uto,Nica:2016elr,Nica:2019ykt,Reich:2012ouk,Browne:2014ukl,Browne:2014gwf,Baglin:2000ong,Baglin:2008hsa}. Except for the cases of ${}^{99}$Tc and ${}^{183}$W noted above, we included all levels in a certain rotational band according to the adopted level determination in the NNDC.

\subsection{Stable LEC Extraction Across EFT Orders and Additional Data}
In this subsection we show that lower-order LECs extracted for the selected rotational bands are stable across EFT orders and with the addition of high-energy data, provided that we account for enough omitted terms when treating the theory error. For most nuclei including omitted terms up to $k_{max}=10$, i.e., accounting for six omitted terms at \NNNNLO, was enough to stabilize the extraction of the LECs.
\begin{figure}[t!]
    \centering
    \includegraphics[width=\columnwidth]{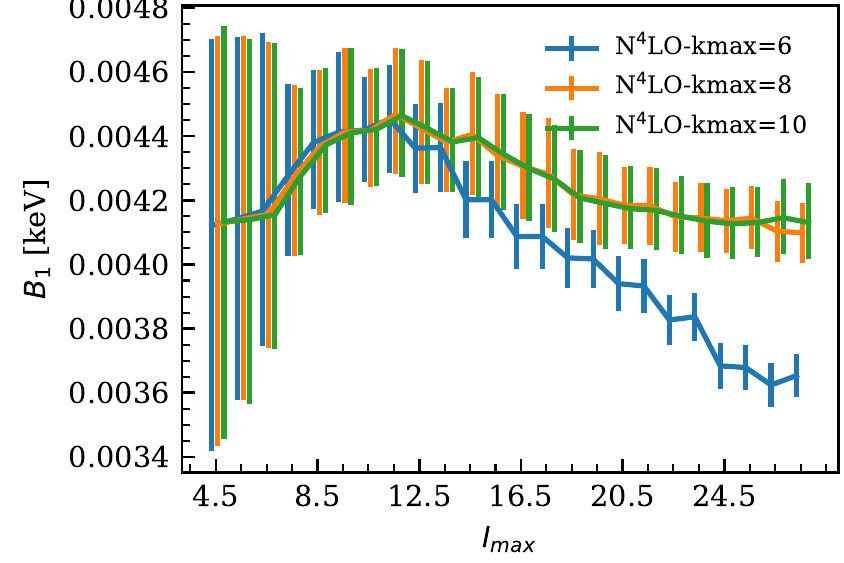}
    \includegraphics[width=\columnwidth]{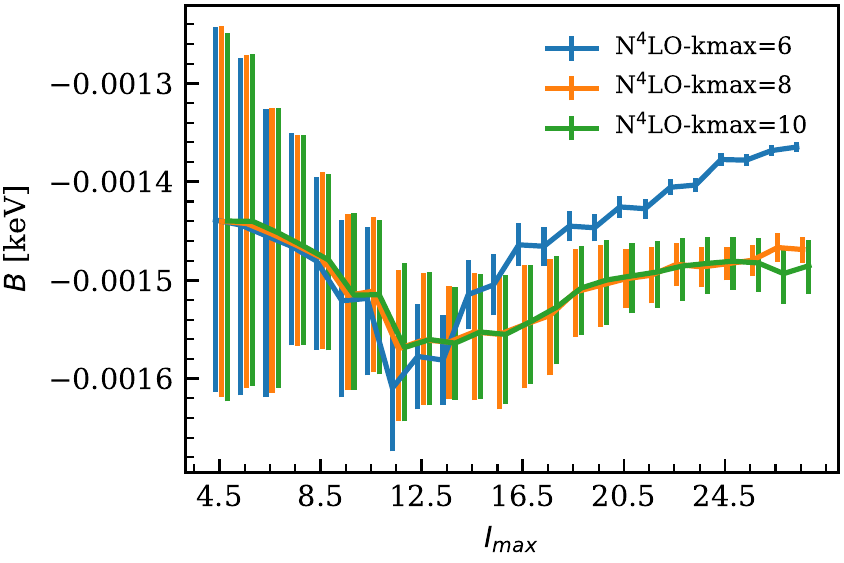}
    \caption{Posteriors for $B_1$ and $B$ describing $^{239}$Pu a function of $I_{max}$ for different values of $k_{max}$. The solid line connects the median values and the error bands encompass the 16th and 84th percentiles in the marginalized distribution.}
    \label{fig:wes_B_above_errors}
\end{figure}

As an example, we show the stability of the extracted LEC, $A_1$, across number of levels included at different EFT orders in Fig.~\ref{fig:wes_A1_above_orders}. In this figure, $I_{max}$ is the spin of the highest-energy level included in a particular analysis. The central values of the resulting posteriors are consistent with each other within 68\% credible intervals, shown as error bars in the figure. Adding more levels to the analysis narrows the posteriors for the LECs up to a certain $I_{max}$, after which the widths of these distributions saturate. 
Fig.~\ref{fig:wes_A1_above_orders} also demonstrates striking agreement between the  distributions obtained at low and high EFT orders: they are almost identical as long as omitted terms up to the same $k_{max}$ are accounted for in both analyses.

The importance of including more than one omitted term in the theory error estimate is evident in Fig.~\ref{fig:wes_B_above_errors}.
The top and bottom panels of the figure show the way that posteriors for $B_1$ and $B$ evolve as $I_{max}$ increases. This is done using three  error models that include different numbers of omitted terms. These results show that including more omitted terms in the  model of the theory error removes the drifting and staggering of the central values.

For both cases the distributions at $k_{max}=10$ agree within errors as we go higher in $I_{max}$. The narrowing of the distribution as we go higher in $I_{max}$ is clearly seen in those two figures. In addition to having less data, the broadening of the error bands at low $I_{max}$ comes from the fact that including less levels in the analysis leads to highly correlated LECs. This allows the numerically larger errors on the lower-order LECs to contribute to the errors on the higher-order LECs, thereby enhancing them.
\begin{figure}[t!]
    \centering
    \includegraphics[width=\columnwidth]{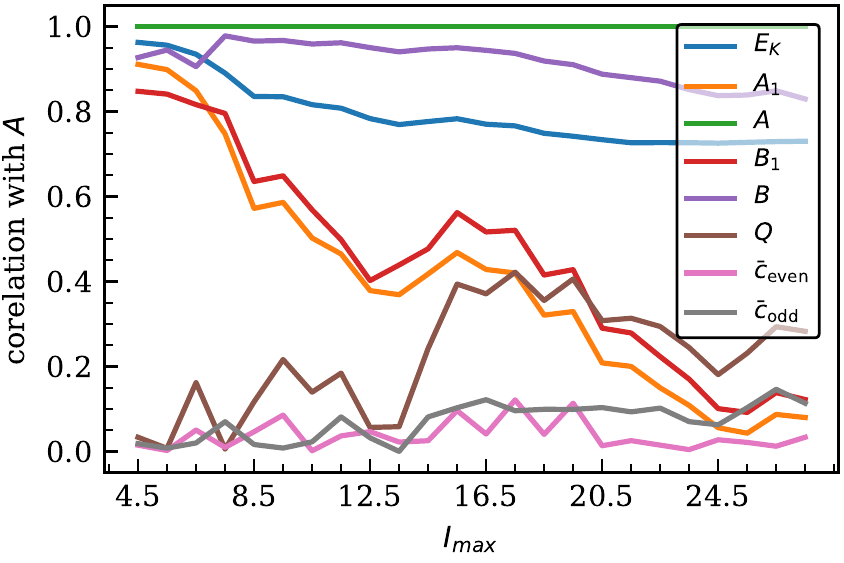}
    \caption{Correlations between LECs and error-model parameters as a function of $I_{max}$, resulting from the analysis on the lowest $K=1/2$ rotational band in $^{239}$Pu at \NNNNLO with $k_{max}=10$.}
    \label{fig:wes_corr}
\end{figure}
\begin{figure}[t!]
    \centering
    \includegraphics[width=\columnwidth]{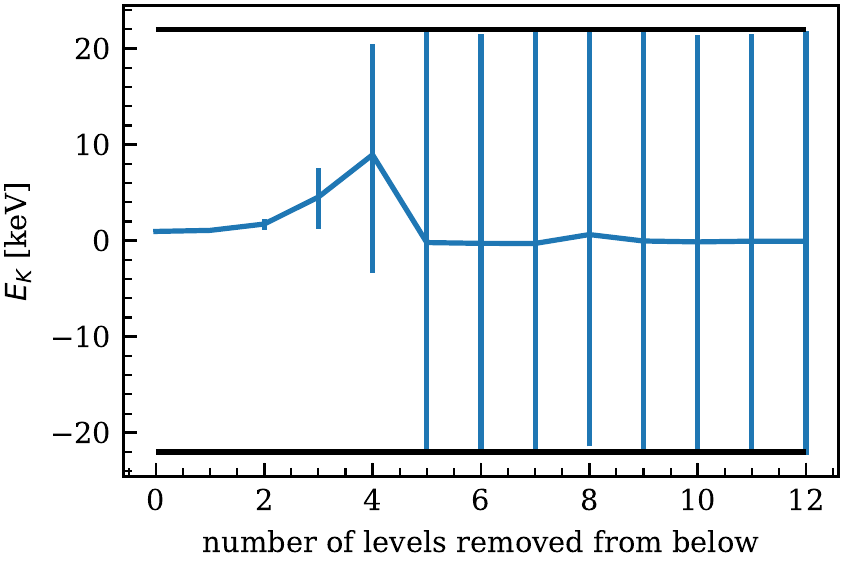}
    \caption{The distribution of $E_K$ for $^{169}$Er at \NNNNLO and $k_{max}=10$ as we successively remove the lowest energy levels from the data set $D$. The solid blue line connects the median values and the error bands encompass uncertainties between on the 16th and 84th percentiles of the samples in the marginalized distribution. The solid black lines show to size of the standard deviation set with the Gaussian prior on $E_K$.}
    \label{fig:wes_EK_below}
\end{figure}

In Fig.~\ref{fig:wes_corr} we show the decrease in the correlations between the LECs as $I_{max}$ increases. The high correlation between the LECs at low $I_{max}$ occurs because these analyses do not include enough data to constrain all LECs independently. Furthermore, the high correlation between the LECs at low $I_{max}$ also results in an unreliable extraction of the expansion parameter $Q$. This comes from the fact that at low energies the theory truncation error is very small compared to the experimental error. Indeed, adding more terms to our EFT error model (i. e., increasing $k_{max}$) leads to higher correlation between the LECs at low $I_{max}$. Thus, the number of levels required to reliably extract $Q$ increases with increasing $k_{max}$.

Starting instead at the low-$I$ end of the data: when we progressively remove the lowest-energy levels from the data set $D$ used to construct the likelihood we rapidly lose the ability to reliably extract the LECs. Figure \ref{fig:wes_EK_below} shows that the distribution for $E_K$ starts narrow and broadens as we remove levels from below. When we remove the six lowest energy levels the distribution of $E_K$ is exactly the same as the prior distribution: the likelihood is making no contribution to the $E_K$ posterior.

The previous results were nearly the same for all cases considered in this work. However, even for $k_{max}=10$, staggering and shifting of the LECs remains sizable for the $K=1/2$ bands in $^{183}$W, $^{167}$Tm and $^{235}$U. In $^{183}$W and $^{167}$Tm, these effects could be attributed to large expansion parameters, as they translate to large omitted contributions to the energies of the rotational levels. In $^{235}$U, the fermionic matrix elements could be larger than naively expected, causing the systematic expansion of the EFT to be questionable as discussed in Ref.~\cite{Alnamlah:2020}.

For $K=3/2$ bands, we were able to extract stable LECs from the $^{159}$Dy analysis by setting $I_{max}=15.5$. This extraction required us to consider omitted terms up to $k_{max}=12$. This is because the spin at which the EFT breaks in this nucleus is $I_{br}\approx 15.5$. (This, then, is the third case in which we do not use all the NNDC energy-level data available on a particular band.) $^{157}$Gd is stable across orders and $I_{max}$, while $^{155}$Gd exhibits shifting and staggering due to a larger expansion parameter, $Q \approx 0.07$.
\begin{table*}[t!]
    \centering
    \renewcommand{\arraystretch}{1.5}
    \begin{tabular}{|c|c|c|c|c|c|c|c|c|}
    \hline
Nucleus&$E_K$ [keV]&$A_1$ [keV]&$ A$ [keV]&$(B_1,A_3)$ [keV]&$B$ [keV]&$Q$&$\bar{c}_{\rm even}$&$\bar{c}_{\rm odd}$\\ \hline 
$^{99}$Tc&$147.3111_{-0.0142}^{0.0143}$&$70.1912_{-0.0103}^{0.0103}$&$82.2577_{-0.0086}^{0.0086}$&$-7.5851_{-0.0022}^{0.0022}$&$-2.99165_{-0.00079}^{0.00079}$&$0.204_{-0.016}^{0.021}$&$1.87_{-0.53}^{0.43}$&$1.35_{-0.39}^{0.54}$\\ \hline
$^{155}$Gd&$-45.2063_{-0.0025}^{0.0025}$&-&$12.0184_{-0.0009}^{0.0009}$&$-0.0081_{-0.0001}^{0.0001}$&$0.00571_{-0.00008}^{0.00008}$&$0.067_{-0.006}^{0.007}$&$3.54_{-1.11}^{2.00}$&$2.01_{-0.94}^{1.99}$\\ \hline
$^{157}$Gd&$-41.3241_{-0.0097}^{0.0093}$&-&$11.0240_{-0.0032}^{0.0034}$&$-0.0094_{-0.0002}^{0.0002}$&$-0.00500_{-0.00025}^{0.00024}$&$0.057_{-0.006}^{0.006}$&$5.23_{-2.33}^{5.18}$&$0.55_{-0.15}^{0.28}$\\ \hline
$^{159}$Dy&$-42.7714_{-0.0093}^{0.0092}$&-&$11.4076_{-0.0030}^{0.0030}$&$-0.0060_{-0.0003}^{0.0003}$&$-0.00307_{-0.00020}^{0.00020}$&$0.063_{-0.004}^{0.004}$&$3.52_{-1.16}^{2.22}$&$0.95_{-0.33}^{0.64}$\\ \hline
$^{167}$Er&$207.2088_{-0.0053}^{0.0053}$&$7.8383_{-0.0028}^{0.0027}$&$11.2398_{-0.0016}^{0.0016}$&$-0.0063_{-0.0003}^{0.0003}$&$-0.00820_{-0.00011}^{0.00010}$&$0.047_{-0.004}^{0.005}$&$4.04_{-1.44}^{2.63}$&$0.83_{-0.26}^{0.49}$\\ \hline
$^{169}$Er&$0.9673_{-0.0197}^{0.0195}$&$9.7774_{-0.0100}^{0.0102}$&$11.7625_{-0.0053}^{0.0058}$&$-0.0064_{-0.0009}^{0.0008}$&$-0.00309_{-0.00023}^{0.00016}$&$0.031_{-0.005}^{0.006}$&$5.57_{-2.51}^{5.31}$&$0.77_{-0.23}^{0.42}$\\ \hline
$^{167}$Tm&$-18.4635_{-0.0155}^{0.0156}$&$-9.1267_{-0.0057}^{0.0056}$&$12.5095_{-0.0022}^{0.0022}$&$0.0431_{-0.0003}^{0.0003}$&$-0.00906_{-0.00008}^{0.00008}$&$0.056_{-0.005}^{0.005}$&$2.16_{-0.76}^{1.37}$&$1.06_{-0.34}^{0.62}$\\ \hline
$^{169}$Tm&$-19.0532_{-0.0011}^{0.0011}$&$-9.7204_{-0.0008}^{0.0008}$&$12.4738_{-0.0007}^{0.0006}$&$0.0264_{-0.0002}^{0.0002}$&$-0.00497_{-0.00006}^{0.00007}$&$0.045_{-0.007}^{0.008}$&$2.59_{-1.12}^{2.30}$&$0.94_{-0.30}^{0.60}$\\ \hline
$^{183}$W&$-6.8750_{-0.0008}^{0.0008}$&$2.7512_{-0.0005}^{0.0005}$&$12.7754_{-0.0004}^{0.0004}$&$-0.0436_{-0.0001}^{0.0001}$&$0.01977_{-0.00003}^{0.00003}$&$0.076_{-0.006}^{0.007}$&$3.80_{-1.49}^{2.86}$&$0.67_{-0.20}^{0.35}$\\ \hline
$^{235}$U&$-6.1882_{-0.0008}^{0.0008}$&$-1.7298_{-0.0007}^{0.0007}$&$6.0508_{-0.0002}^{0.0002}$&$0.0025_{-0.0001}^{0.0001}$&$-0.00249_{-0.00001}^{0.00001}$&$0.040_{-0.003}^{0.004}$&$3.85_{-1.28}^{2.36}$&$0.82_{-0.27}^{0.53}$\\ \hline
$^{239}$Pu&$-8.3578_{-0.0019}^{0.0019}$&$-3.6559_{-0.0011}^{0.0011}$&$6.2726_{-0.0004}^{0.0004}$&$0.0041_{-0.0001}^{0.0001}$&$-0.00149_{-0.00003}^{0.00003}$&$0.029_{-0.003}^{0.004}$&$5.72_{-2.22}^{4.18}$&$0.93_{-0.31}^{0.61}$\\ \hline

    \end{tabular}
    \renewcommand{\arraystretch}{1}
    \caption{The median value of the LECs and the error-model parameters at \NNNNLO for the nuclei considered in this work. The uncertainties encompass the 16th and 84th percentiles of the samples in the marginalized distributions. $K=3/2$ rotational bands do not have a parameter $A_1$ and the parameters ($B_1,A_3$) refer to $K=1/2$ and $K=3/2$ bands respectively.}
    \label{tab:LECs}
\end{table*}

The values of the LECs and the error-model parameters at \NNNNLO for the nuclei considered in this work are given in Tabele \ref{tab:LECs}.

\subsection{Prior Sensitivity}
In addition to using the scaled-inverse-$\chi^2$ distribution as a prior for $\bar{c}_{even}$ and $\bar{c}_{odd}$ we tried truncated Gaussians with mean zero and standard deviations $\sigma=7$ and $\sigma=3$ respectively for all cases. These truncated Gaussian priors allow for smaller values of the characteristic sizes. 
But the standard deviations were chosen to still allow values for $\bar{c}_{odd}$ and $\bar{c}_{even}$ larger than those resulting from scaled-inverse-$\chi^2$ priors with $\nu=1$ and $\tau^2=1$.

The change in prior for $\bar{c}_{odd}$ and $\bar{c}_{even}$ does not significantly change the posteriors for the LECs: the corresponding central values differ by less than 1\%, and are consistent with each other within the 68\% credible intervals. 
Central values of the posteriors for $Q$ differ by less than 15\%, and were similarly consistent. 

The strongest dependence on the prior is that exhibited by the posteriors for $\bar{c}_{odd}$ and $\bar{c}_{even}$: the central values differ in some cases by more than 50\%. However, even these values are consistent with each other within 68\% credible intervals, since the posteriors for the characteristic sizes are broad.

The changes in the posteriors of $Q$ on one hand and $\bar{c}_{odd}$ \& $\bar{c}_{even}$ on the other are anticorrelated. We only care about combinations of them to set the size of the theory error and the expected size of the LECs. Thus, the dependence of the theory error and the expected size of the LECs on the prior for the characteristic sizes is less profound. The difference in sizes of the theory error resulting from the chosen priors is less than 20\% for all cases except $^{157}$Gd, where the difference is about 40\%.

For all results that follow we used the scaled-inverse-$\chi^2$ distribution with $\nu=1$ and $\tau^2=1$ as the prior for both $\bar{c}_{odd}$ and $\bar{c}_{even}$, in keeping with the naturalness assumption.
\subsection{Posterior Predictive Distributions}
Figure \ref{fig:Er_ppd,U_ppd} shows the PPDs of the energy residuals as a function of the spin $I$ for two cases considered in this work. These distributions are calculated using Eq.~(\ref{PPD}). 
In each figure, translucent blue lines connect energy residuals resulting from different LECs sets sampled from the posterior distribution in Eq.~\eqref{eq:master}. The solid black line represents the median of the PPD, and the dashed lines encompass the region between the 16th and 84th percentiles.
The dark and light red bands show the truncation error and the experimental error added in quadrature at 68\% and 95\% credible levels respectively. To calculate the truncation  error, we consider a theory error that accounts for six omitted terms. They are added in quadrature, just as they are in the likelihood defined in Sec.~\ref{sec:Bayesianmodel}. This calculation was done using the median values of $Q, \bar{c}_{even}$, and $\bar{c}_{odd}$. The dependence of the size of the theory error on the prior on $\bar{c}_{even}$ and $\bar{c}_{odd}$ is small in these cases: the theory error changes by about 10\% when the prior is changed.
\begin{figure}[t!]
    \centering
    \includegraphics[width=\columnwidth]{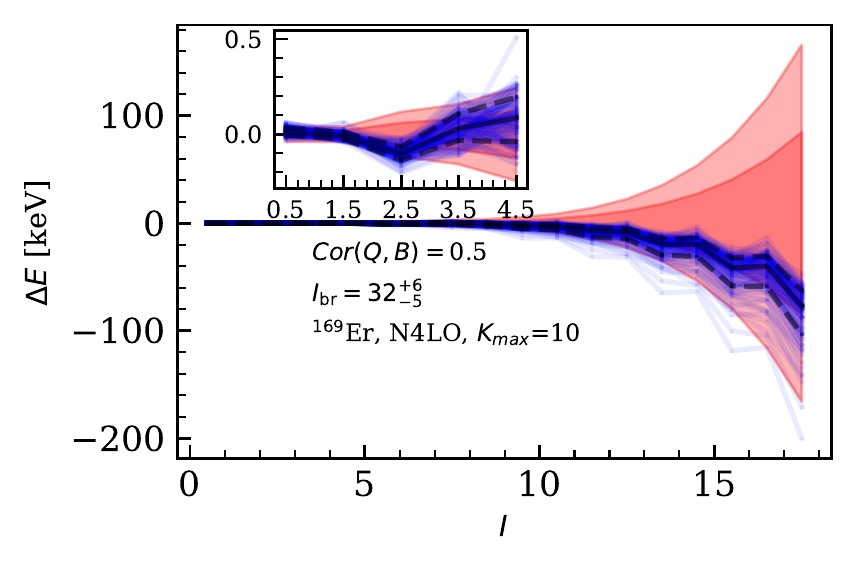}
    \includegraphics[width=\columnwidth]{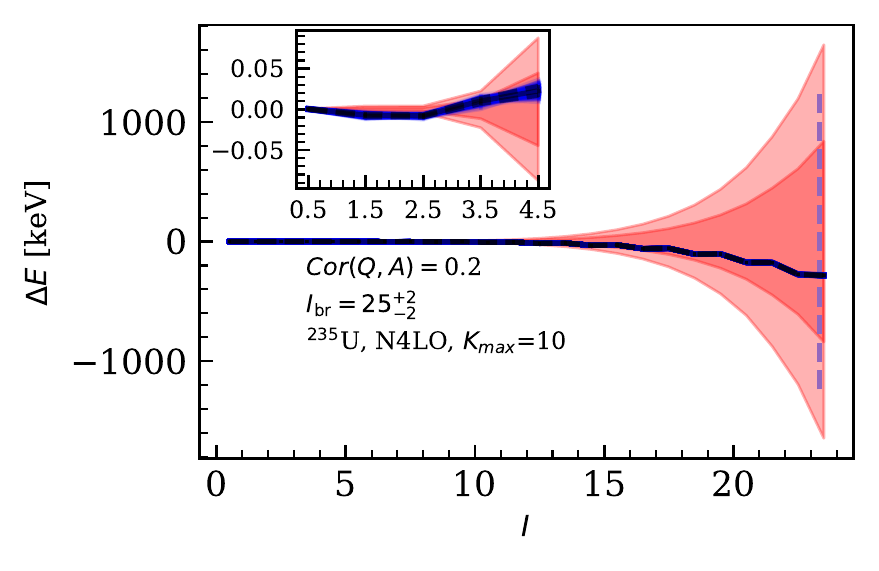}
    \caption{The posterior predictive distribution for energy-level residuals at \NNNNLO and $k_{max}=10$ in $^{169}$Er and $I_{max}=17.5$ (top panel) and in $^{235}$U and $I_{max}=23.5$ (bottom panel). The dark and light red bands show the truncation error plus the experimental error at 68\% and 95\% credible levels respectively. The lighter blue lines connect the energy residuals calculated from the distribution of the LECs. The solid black line represents the median of the distribution and the dashed lines indicate the 16th and 84th percentiles. The correlation shown on the plot is the highest correlation between any LEC and any error-model parameter. $I_{br}$ was determined from the distribution of $Q$. The dashed purple line shows the lower limit of $I_{br}$. The insert on the plot shows the residuals on the first 5 levels with an altered y-axis scale.} 
    \label{fig:Er_ppd,U_ppd}
\end{figure}

The correlation coefficient written in the legend in Fig. \ref{fig:Er_ppd,U_ppd} is the largest between any LEC and any error-model parameter for the shown analysis. When this value is small, the truncation error and the propagated LEC error could in principle be added together in quadrature.
 
In viewing Fig.~\ref{fig:Er_ppd,U_ppd} it is important to remember that 
the truncation error on the energy residuals is highly correlated across levels. This comes from the high correlation between levels when building the correlation matrix that goes into the likelihood. This correlation also flows into a correlation between levels in the PPD of the energies. A correlation plot between two energy levels, like the ones in Fig. \ref{fig:Er_2d_ppd,U_2d_ppd}, gives a 2D cut of this multi-dimensional correlation.
  \begin{figure}[t!]
    \centering
    \includegraphics[width=\columnwidth]{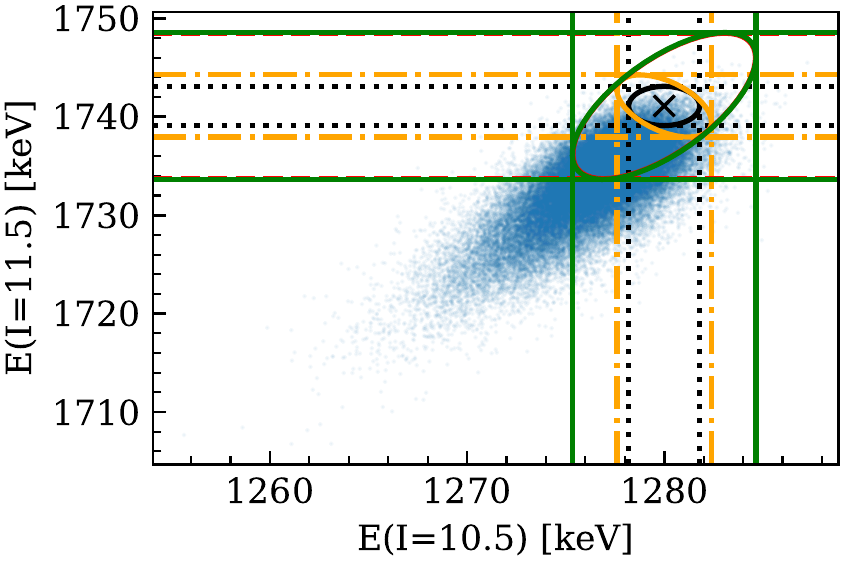}
    \includegraphics[width=\columnwidth]{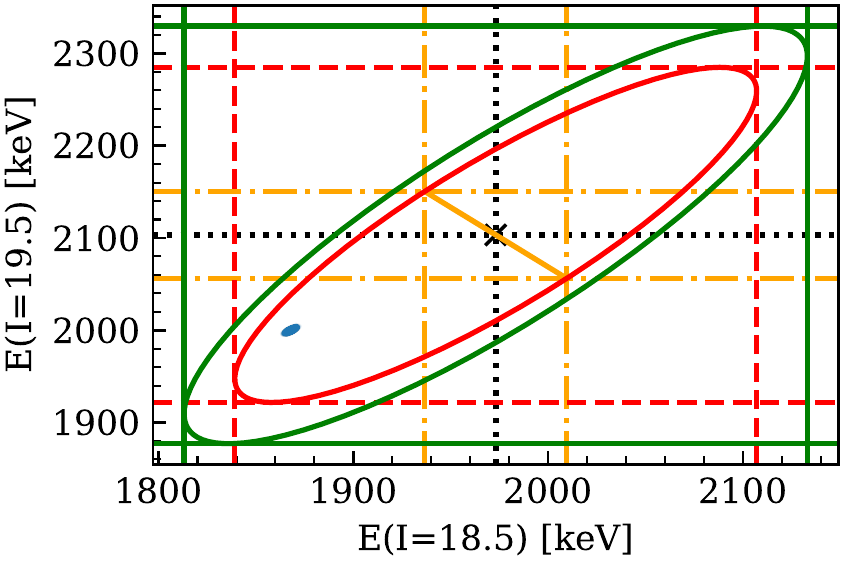}
    \caption{A 2D cut of the posterior predictive distribution at \NNNNLO and $k_{max}=10$ for $^{169}$Er and $I_{max}=17.5$ (top panel) and in $^{235}$U and $I_{max}=23.5$ (bottom panel). The blue dots show the energies calculated from the distribution of the LECs. The black cross shows the experimental value and the black lines and black ellipse shows the corresponding experimental uncertainty. The remaining ellipses and lines show the truncation error and the experimental error added in quadrature. (All the ellipses are centered at the experimental value.) The orange, red and green account for 1, 2 and 6 omitted terms in the theory error respectively. (In the top panel the red ellipse is completely covered by the green ellipse.)}
    \label{fig:Er_2d_ppd,U_2d_ppd}
\end{figure}

In both panels we see the importance of accounting for more than one omitted term in the theory error. This is clearly shown in the reverse in the direction of the correlation from a negative to a positive correlation when going from the orange ellipse to the red ellipse. The orange ellipse is obtained when we account for only one omitted term, while the red ellipse includes the effect of two omitted terms. After accounting for six omitted terms the green ellipse is obtained and the 68\% ellipse in principle expands. This is more clearly seen when we go to high-energy levels plotted in the lower panel in Fig. \ref{fig:Er_2d_ppd,U_2d_ppd}. Note also that for lower-energy levels the correlation is smaller since the experimental error dominates over the truncation error, and we assumed that the experimental errors are not correlated across energy levels. 

\subsection{Model Checking}
\begin{figure}[h]
    \centering
    \includegraphics[width=\columnwidth]{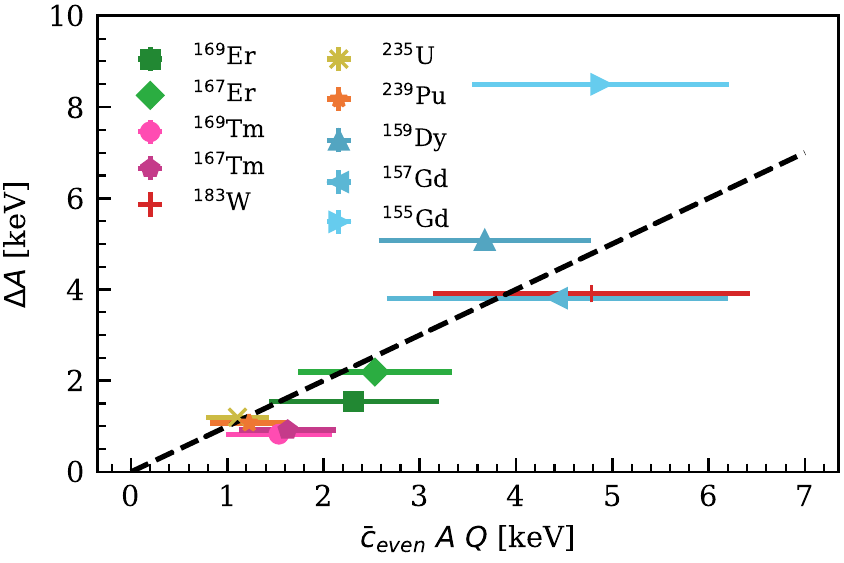}
    \caption{The size of the \NNLO LEC, $\Delta A$, (on the y-axis) compared to its expected size from the EFT power counting (on the x-axis). Error bands on the LEC distribution are small and can not been seen on the plot. The error bands on the x-axis encompass the 16th and 84th percentiles. Different nuclei are labeled in the legend of the plot. The blue colored points are results for rotational bands with bandheads $K=3/2$, all the others are $K=1/2$ bands. }
    \label{fig:A_vs_Q}
\end{figure}

\begin{figure}[h]
    \centering
    \includegraphics[width=\columnwidth]{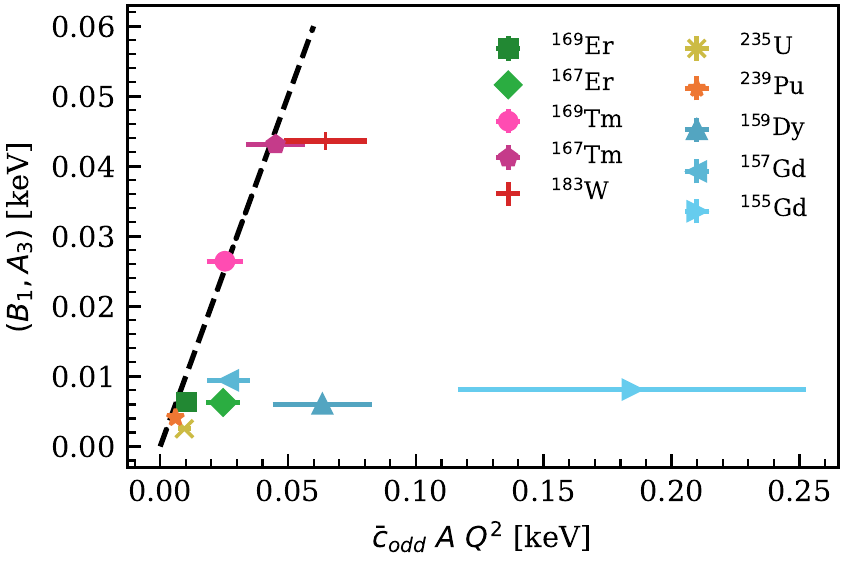}
    \caption{The size of the \NNNLO LEC, $B_1$ for $K=1/2$ bands and $A_3$ for $K=3/2$ bands, (on the y-axis) compared to its expected size from the EFT power counting (on the x-axis). Error bands on the LEC distribution are small and can not been seen on the plot. The error bands on the x-axis encompass uncertainties between on the 16th and 84th percentiles. Different nuclei are labeled in the legend of the plot. The yellow colored points are results for rotational bands with bandheads $K=3/2$, all the others are $K=1/2$ bands.}
    \label{fig:B1_vs_Q}
\end{figure}

\begin{figure}[h]
    \centering
    \includegraphics[width=\columnwidth]{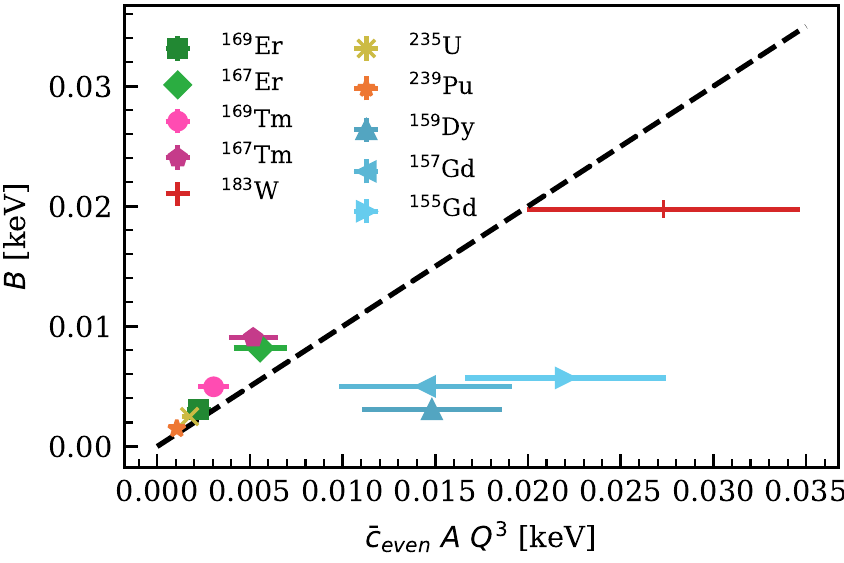}
    \caption{The size of the \NNNNLO LEC, $B$, (on the y-axis) compared to its expected size from the EFT power counting (on the x-axis). Error bands on the LEC distribution are small and can not been seen on the plot. The error bands on the x-axis encompass uncertainties between on the 16th and 84th percentiles. Different nuclei are labeled in the legend of the plot. The yellow colored points are results for rotational bands with bandheads $K=3/2$, all the others are $K=1/2$ bands.}
    \label{fig:B_vs_Q}
\end{figure}

In Figs. \ref{fig:A_vs_Q}, \ref{fig:B1_vs_Q} and \ref{fig:B_vs_Q} we compare the marginalized posterior distributions of the LECs, on the y-axis, with their expected sizes from the EFT power counting, on the x-axis. Since we also extract the theory error parameters from the sampler and they are highly correlated among themselves, we calculate the expected size from the distributions of the error model parameters using Eq. (\ref{PPD}). We notice that the error on the distribution of the LECs is very small compared to the error on the expected sizes that comes from the distribution of the theory error parameters. 

As these graphs are model-checking graphs, and since the estimates of LEC sizes plotted on the x-axis are meant as order-of-magnitude estimates, we do not expect perfect linear correlations. 
Nevertheless, Fig.~\ref{fig:A_vs_Q} shows that, for all $K=1/2$ bands considered, the LEC $\Delta A$ agrees with its expected size within error bands. This result is surprisingly better than expected. In contrast, the size of $\Delta A$ for $K=3/2$ bands is larger than expected, especially for $^{155}$Gd (see blue points in Fig.~\ref{fig:A_vs_Q}). There are two factors that could contribute to this. First, the $K=3/2$ bands have larger fermionic matrix elements. This could hinder the systematic expansion of the EFT. Second, the $K=3/2$ bands have relatively larger expansion parameters, see Fig. \ref{fig:Q_vs_Q}. 

The same discussion applies to the results in Figs. \ref{fig:B1_vs_Q} and \ref{fig:B_vs_Q}, where we see good agreement between the LECs and their expected sizes for $K=1/2$ bands. The disagreement with power-counting estimates for $K=3/2$ bands at \NkLO{3,4} is less of a concern than the one at \NkLO{2} seen in Fig.~\ref{fig:A_vs_Q}, since these higher-order LECs are smaller than their expected sizes. This doesn't undermine the convergence of the EFT expansion.

We also note here that the scale of the x-axis is prior dependent and could change by more than 50\% in some nuclei, depending on the choice of prior on $\bar{c}_{even}$ and $\bar{c}_{odd}$. 
For $^{157}$Gd changing the prior on $\bar{c}_{even}$ and $\bar{c}_{odd}$ to a truncated normal allowed for smaller values of $\bar{c}_{odd}$ and $A_3$ was then equal to the expected size (i. e.,the point for $^{157}$Gd then falls exactly on the line in figure \ref{fig:B1_vs_Q}). This did not happen when the truncated normal is chosen as a prior for the analysis in $^{155}$Gd and $^{159}$Dy; this may occur because there is strong \NkLO{5} energy-level staggering present in the data for these nuclei.

The size of $\bar{c}_{even}$ and $\bar{c}_{odd}$ is constrained by both the sizes of the LECs and the size of the theory error. In a good systematic expansion the tension between those factors on setting the size of the $\bar{c}$'s would be small and one number apiece would suffice to represent the even and odd order corrections. However when the systematic expansion is hindered, as in the case for $K=3/2$ bands due to large fermionic matrix elements, this tension becomes clear. One example of this is seen in Figs. \ref{fig:A_vs_Q} and \ref{fig:B_vs_Q} for $K=3/2$ bands. There $\Delta A$ is large and favors large values of $\bar{c}_{even}$, however, $B$ is small and favors smaller values of $\bar{c}_{even}$. The eventual result is a compromise. This tension may be exacerbated by the truncation error also providing information on the size of the $\bar{c}$'s.
\subsection{Higher than Expected Break-down Scale}
\begin{figure}[t!]
    \centering
    \includegraphics[width=\columnwidth]{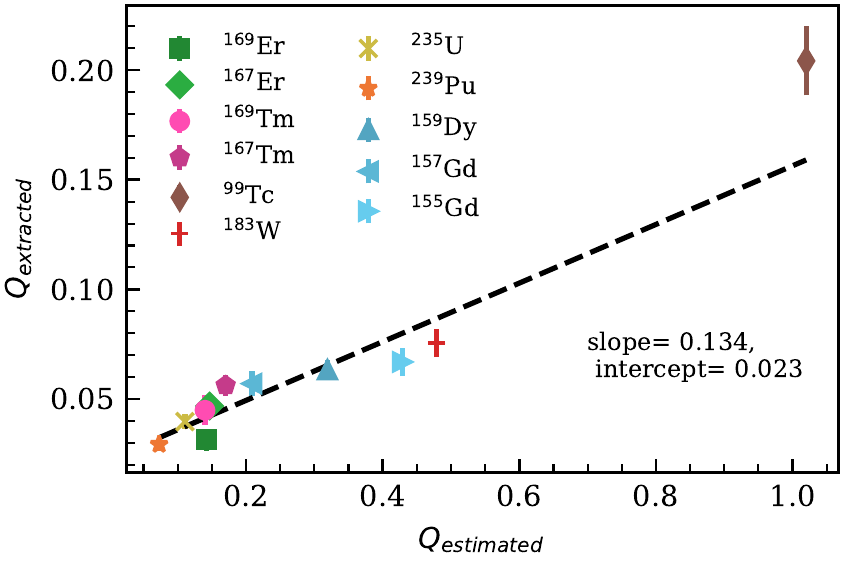}
    \caption{The extracted expansion parameter $Q$ from marginalized posterior distribution of our sampling compared to its naively expected size. The expected size is taken to be the maximum of $E_{\rm rot}/E_{\rm sp}$ and $E_{\rm rot}/E_{\rm vib}$. The dashed black line shows the best linear fit and its parameters are printed on the plot. The blue colored points are results for rotational bands with bandheads $K=3/2$, all the others are $K=1/2$ bands.}
    \label{fig:Q_vs_Q}
\end{figure}

In Fig. \ref{fig:Q_vs_Q} we see a clear correlation between the extracted values of $Q$ and those that are expected based on each nucleus' single-particle and vibrational energy scales, $E_{\rm sp}$ and $E_{\rm vib}$. The expected $Q$ is the larger of $E_{\rm rot}/E_{\rm sp}$ and $E_{\rm rot}/E_{\rm vib}$, while the extracted $Q$ comes from sampling the posterior in Eq. \ref{posterior}. This extracted $Q$ is what actually determines the convergence of the EFT expansion. It is markedly smaller than would be naively expected. The break-down scale of the theory is thus higher than naively expected: our rotational EFT works to much higher $I$ than energy-scale arguments would suggest. This could occur because coupling between the higher rotational states explicitly included in the EFT and the high-energy states that are not explicitly included in our EFT is hindered by the large difference in angular momentum between them.
\section{Conclusions}
\label{sec:conclusion}
We performed a Bayesian analysis to extract the LECs and expansion parameters $Q$ describing the rotational energy levels of diverse odd-mass nuclei within a recently developed EFT.
This analysis corroborates the EFT organization for energy-level formulae which results from the assumed power-counting scheme: the extracted LECs of order $k$ scale as $Q^{k-1}$, i.e., according to EFT expectations.
While our analysis reached this conclusion for both $K=1/2$ and $K=3/2$ rotational bands, the sizes of the LECs describing the latter exhibit larger deviations from their expected values than those describing the former. We attribute this behavior to the size of fermionic matrix elements, assumed to be of order one while organizing energy-level formulae. Since these matrix elements involve the angular momentum of the fermion, $\vec{j}$, we cannot exclude the possibility that the systematic behavior of the EFT is hindered in bands build on top of single-particle orbitals with larger values of $K$. For the $K=3/2$ bands studied in this work, however, this discrepancy does not destroy the systematic improvement of calculated energies up to \NNNNLO, as the sizes of extracted LECs are smaller that expected.

In order to ensure that the extracted values are independent of the EFT order and number of energy levels entering the analysis, we employed a theory error beyond the first-omitted-term approximation, considering omitted terms in the expansion for the energy of rotational levels up to order $k_{max}$. As we increased the number of omitted terms considered in the theory error, the corresponding log likelihood exhibited steeper and steeper peaks. Therefore, the `widths' of the sampled posteriors decrease as $k_{max}$ increases. Considering six omitted terms at \NNNNLO enabled a stable extraction of the LECs and expansion parameters describing the levels of interest. The shapes of posteriors for low-order LECs extracted at this order and those extracted using lower-order energy formulae are, for all practical purposes, identical. On the other hand, the shapes of the posteriors depend strongly on the number of levels informing the model, narrowing as more levels are included. Nevertheless, the 68\% credible intervals of these posteriors possess significant overlap, facilitating reliable LEC extraction.

In addition to the posteriors for the LECs and expansion parameters, our analysis yielded distributions for the characteristic sizes of even and odd $c_n$'s, $\bar{c}_{even}$ and $\bar{c}_{odd}$. The values of $\bar{c}_{odd}$ are typically smaller than those for $\bar{c}_{even}$, in agreement with results from previous studies where the LECs were fitted to the smallest possible data sets. The difference of the characteristic sizes of even and odd LECs has its origin in the physics behind the corresponding contributions to the effective Lagrangian: while odd-order contributions correct the particle-rotor interaction, even-order contributions include terms that depend exclusively on the rotor degrees of freedom, thus correcting the physics of the core to which the particle is coupled. Here this conclusion was reached solely on the basis of experimental data; we assumed equal priors for both characteristic sizes.

Although the distributions of $\bar{c}_{odd}$ and $\bar{c}_{even}$ change depending on the choice of the their priors, that does not significantly change the distributions of the LECs. Altering the priors also does not have a large effect on the size of the theory error, which changes by less than 20\% for nearly all cases. 

These considerations mean that our extractions of the LECs and the theory error parameters in the EFT of rotational bands in odd-mass nuclei are robust under the choice of prior.  The formalism presented here also gives robust results for LECs across orders and as more data is added to the analysis. We conclude that a Bayesian framework that incorporates theory errors in the likelihood offers significant advantages for LEC extraction in EFTs. This methodology has already been used for the extraction of LECs in the NN potential from phase shifts~\cite{Wesolowski:2019} and to constrain parameters of the three-nucleon force~\cite{Wesolowski:2021}. But it is a very general approach which should improve the parameter estimation for LECs in any EFT. 

\begin{acknowledgments}
We thank Dick Furnstahl and Jordan Melendez for useful discussions. We are also grateful for Daniel Odell's significant conceptual and computational assistance.
This work was supported by the US Department of Energy, contract DE-FG02-93ER40756 (IKA, DRP) by the National Science Foundation CSSI program Award OAC-2004601 (DRP), and under the auspices of the US Department of Energy by Lawrence Livermore National Laboratory under Contract DE-AC52-07NA27344 (EACP). IKA acknowledges the support of King Saud University and the Ministry of Education in Saudi Arabia.
\end{acknowledgments}

\end{document}